\documentclass[aps,pre,floats,twocolumn,showpacs,superscriptaddress]{revtex4}

\usepackage{graphicx,epsfig}
\usepackage{times}
\usepackage{graphics,dcolumn,bm,fleqn,epic,eepic,float}
\usepackage{amssymb,amsmath,multirow,rotate,color}
\begin{document}

\title{Complex cooperative networks from evolutionary preferential attachment}
\author{Julia Poncela}

\affiliation{Institute for Biocomputation and Physics of Complex
Systems (BIFI), Universidad of Zaragoza, 50009 Zaragoza, Spain}

\author{Jes\'us G\'omez-Garde\~nes}

\affiliation{Institute for Biocomputation and Physics of Complex
Systems (BIFI), Universidad of Zaragoza, 50009 Zaragoza, Spain}

\affiliation{Scuola Superiore di Catania, Universit\`a di Catania, 
95123 Catania, Italy}

\author{Luis M. Flor\'{\i}a}

\affiliation{Institute for Biocomputation and Physics of Complex
Systems (BIFI), Universidad of Zaragoza, 50009 Zaragoza, Spain}

\affiliation{Departamento de F\'{\i}sica de la Materia Condensada,
  Universidad de Zaragoza, 50009 Zaragoza, Spain}
  
  \author{Angel S\'anchez}

\affiliation{Institute for Biocomputation and Physics of Complex
Systems (BIFI), Universidad of Zaragoza, 50009 Zaragoza, Spain}

\affiliation{Grupo Interdisciplinar de Sistemas Complejos (GISC),
Departamento de Matem\'aticas, Universidad Carlos III de Madrid, 28911 Legan\'es, Spain}

\affiliation{IMDEA Matem\'aticas, Campus de la Universidad Aut\'onoma de Madrid, Facultad de Ciencias, 28049 Madrid}
  
\author{Yamir Moreno}\thanks{E-mail: yamir@unizar.es}

\affiliation{Institute for Biocomputation and Physics of Complex
Systems (BIFI), Universidad of Zaragoza, 50009 Zaragoza, Spain}

\affiliation{Department of Theoretical Physics, University of Zaragoza, 50009 Zaragoza, Spain}

\date{\today}

\begin{abstract}

In spite of its relevance to the origin of complex networks, 
the interplay between form and function and its role during network formation 
remains largely unexplored. 
While recent studies introduce dynamics by considering rewiring 
processes of a pre-existent network, we study network growth and formation
by proposing an evolutionary 
preferential attachment model, its main feature being that the capacity of a node to attract new links depends on a dynamical variable governed in turn by the node interactions. As a specific example, we focus on the problem of the emergence of cooperation by analyzing
the formation of a social network with 
interactions given by the Prisoner's Dilemma. The resulting networks show many features of real systems, such as scale-free degree distributions, cooperative behavior and hierarchical clustering. 
Interestingly, results such as the cooperators being located mostly on nodes of intermediate 
degree are very different from the observations of cooperative behavior on static networks. 
The evolutionary preferential attachment
mechanism points to an evolutionary origin of scale-free networks and may help understand
similar feedback problems in the dynamics of complex networks by appropriately choosing the game 
describing the interaction of nodes. 
\end{abstract}

\pacs{05.45.Xt, 89.75.Fb}
\maketitle

\section{Introduction}

In the last few years, it has been increasingly realized that there are many situations which
are not well described by well-mixed (mean-field) models, lattices and uniformly distributed spatial models. This is 
the case with the majority of the so called complex systems, that are better characterized by what is 
generally known as complex networks \cite{siam,PhysRep}. 
In many of these networks, the distribution of the number
of interactions, $k$, that an individual shares with the rest of the elements of the system,
$P(k)$, is found to
follow a power-law, $P(k)\sim k^{-\gamma}$, with $2<\gamma<3$ in most cases. The ubiquity in Nature of these so-called scale-free (SF) networks has led scientists to propose many models aimed at reproducing the SF degree distribution 
\cite{siam,PhysRep}. Most of the existing approaches are based on growth rules that depend on the
instantaneous topological properties of the network and therefore neglect the connection of the structural
evolution and the particular function of the network. However, 
accumulated evidence suggests, moreover, that form follows function \cite{guimera} and that the formation of the network is also related to the dynamical states of its components through a feedback mechanism that shapes its structure.

On the other hand, a paradigmatic case study of the structure
and dynamics of complex systems is that of social networks. In these systems, it is
particularly relevant to understand how cooperative behavior emerges. The mathematical approach to model the (cooperative versus
defective) interactions is usually tackled under the general framework of
evolutionary game theory through diverse social dilemmas. In the general case
it is the individual benefit rather than the overall welfare what drives the
system evolution. The emergence of cooperation in natural and social
systems has been the subject of intense research recently \cite{sp05,lhn05,gcfm07,pgfm07,ohln06,ezcs05,spl06,Nowak1,jlcs07,las08}. These works are based either on the assumption of an underlying, given static network (or two static, separate networks for interaction and imitation \cite{onp07}) or a coevolution and rewiring starting from a fully developed network 
that already includes all the participating elements. The results show that if the well-mixed population hypothesis is abandoned, so that
individuals only interact with their neighbors, cooperation is promoted on 
heterogeneous networks, specifically on SF networks. However, the main questions remain unanswered: Are cooperative behavior and structural properties of networks related or linked 
in any way? If so, how? Moreover, if SF networks are best suited to support cooperation, then, where did they come from? What are the mechanisms that shape the structure of the system? 

In this paper we analyze the growth and formation of complex networks by
coupling the network formation rules to the dynamical states of the elements of the system. With 
the problem of the emergence of cooperation as a specific application in mind, we consider that the nodes of the network are individuals involved in a social dilemma and that newcomers are preferentially linked to nodes with high fitness, the latter being proportional to the payoffs obtained in the game. In this way, the fitness of an element is not imposed as an external constraint \cite{bb01,cal02}, but rather it is the result of the dynamical evolution of the system. At the same time, the network is not exogenously imposed as a starting point but instead it grows from a small seed and acquires its structure during its formation process. The main result of this interplay is the formation of  homogeneous and heterogeneous networks that share a number of topological features with real world networks such as a high clustering and degree-degree correlations. Remarkably, the set of nodes sustaining the 
observed aggregate behavior is very different from that arising in a complex but otherwise static network. 
As a particular but most relevant conclusion, 
we find that the mechanism we propose  not only explains why heterogeneous networks are tailored to sustain cooperation, but also provides an evolutionary mechanism for their origin.

\section{Evolutionary Preferential Attachment model}

Our model naturally incorporates an intrinsic feedback between dynamics and topology. The growth of the network starts at time $t=0$ with a core of $m_{0}$ fully connected nodes. New elements are incorporated to the network and
attached to $m$ existing nodes with a probability that depends on the dynamics of each node. In particular, we consider that 
the dynamics is dictated by the Prisoner's Dilemma (PD) game. In this
two-players game, every node initially adopts with the same probability \cite{note3} one of
the two available strategies, cooperation $C$ or defection $D$. At
equally spaced time intervals (denoted by $\tau_{D}$) each node $i$ of the network plays with its $k_{i}(t)$ neighbors and 
the obtained payoffs are considered to be the measure of its evolutionary fitness,
$f_i(t)$.  There are three possible situations for each link in the network:
{\em (i)} if two cooperators meet both receive $R$, when {\em (ii)} two
defectors play both receive $P$, while {\em (iii)} if a cooperator and
a defector compite the former receives $S$ and the latter obtains $T$. The four
payoffs are ordered as $T=b>R=1>P=S=0$. After playing, every node $i$ compares its
evolutionary fitness (payoff) with that corresponding to a randomly chosen neighbor
$j$. If $f_{j}(t)>f_{i}(t)$ node $i$ adopts the strategy of player $j$ with
probability \cite{note2}
\begin{equation}
P_{i}=\frac{f_{j}(t)-f_{i}(t)}{b\cdot{\text {max}}\left[k_{i}(t),k_{j}(t)\right]}\;.
\label{replicator}
\end{equation}

The growth of the network proceeds by adding a new node with $m$ links to the preexisting
ones at equally spaced time intervals (denoted
by $\tau_{T}$). The probability that any node $i$ in the network receives one of the $m$ new links is
\begin{equation}
\Pi_{i}(t)=\frac{1-\epsilon+\epsilon
f_{i}(t)}{\sum_{j=1}^{N(t)}(1-\epsilon+\epsilon f_{j}(t))}\;,
\label{Pattach}
\end{equation}
where $N(t)$ is the size of the network at time $t$. The parameter $\epsilon\in [0,1)$ thus controls the weight of the $f_{i}(t)$'s during the growth of the network. When $\epsilon>0$, nodes with $f_i(t)\neq 0$ are preferentially chosen. 

\begin{figure}
\begin{center}
\epsfig{file=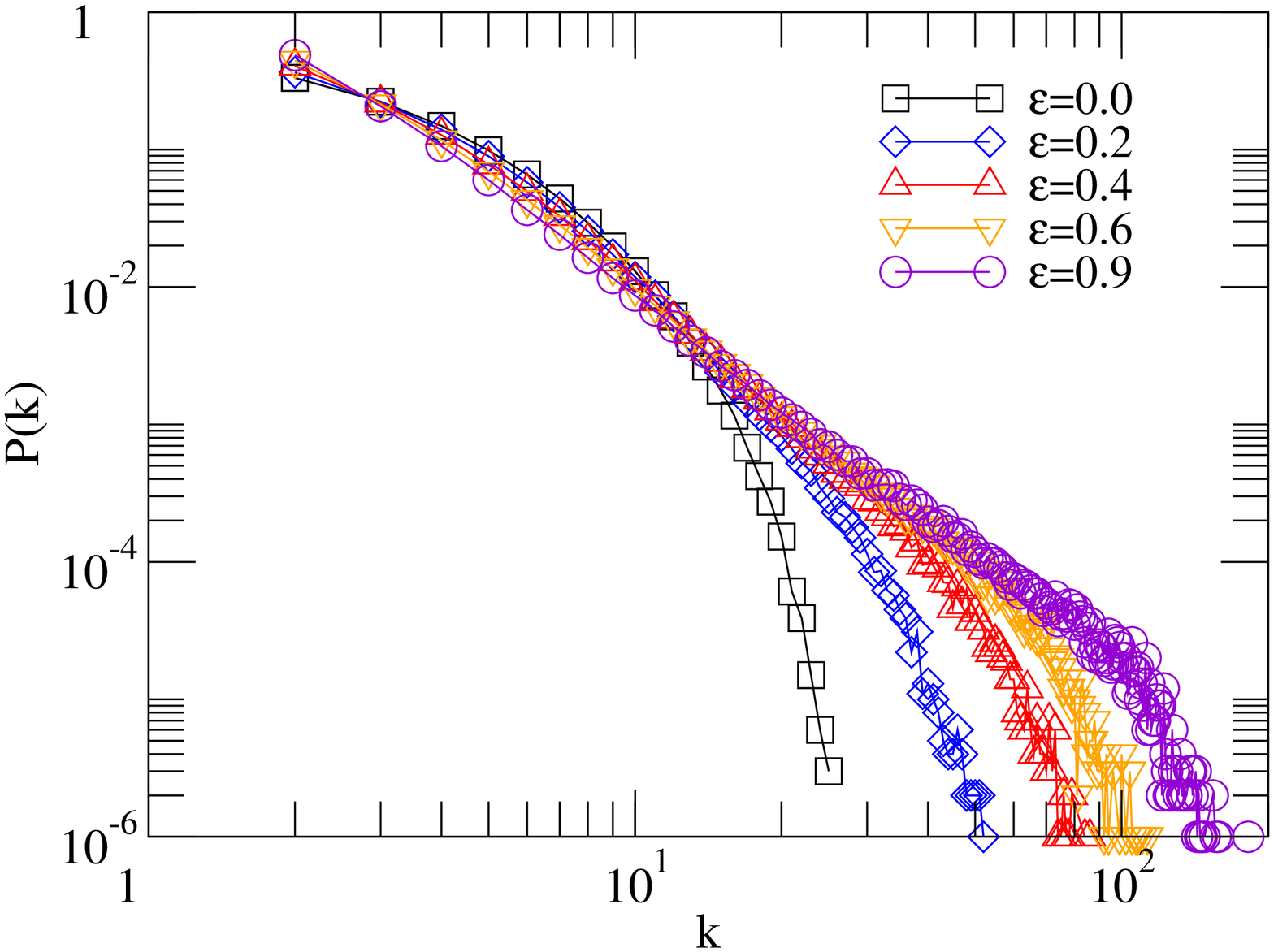,width=0.8\columnwidth,angle=-90,clip=1}
\epsfig{file=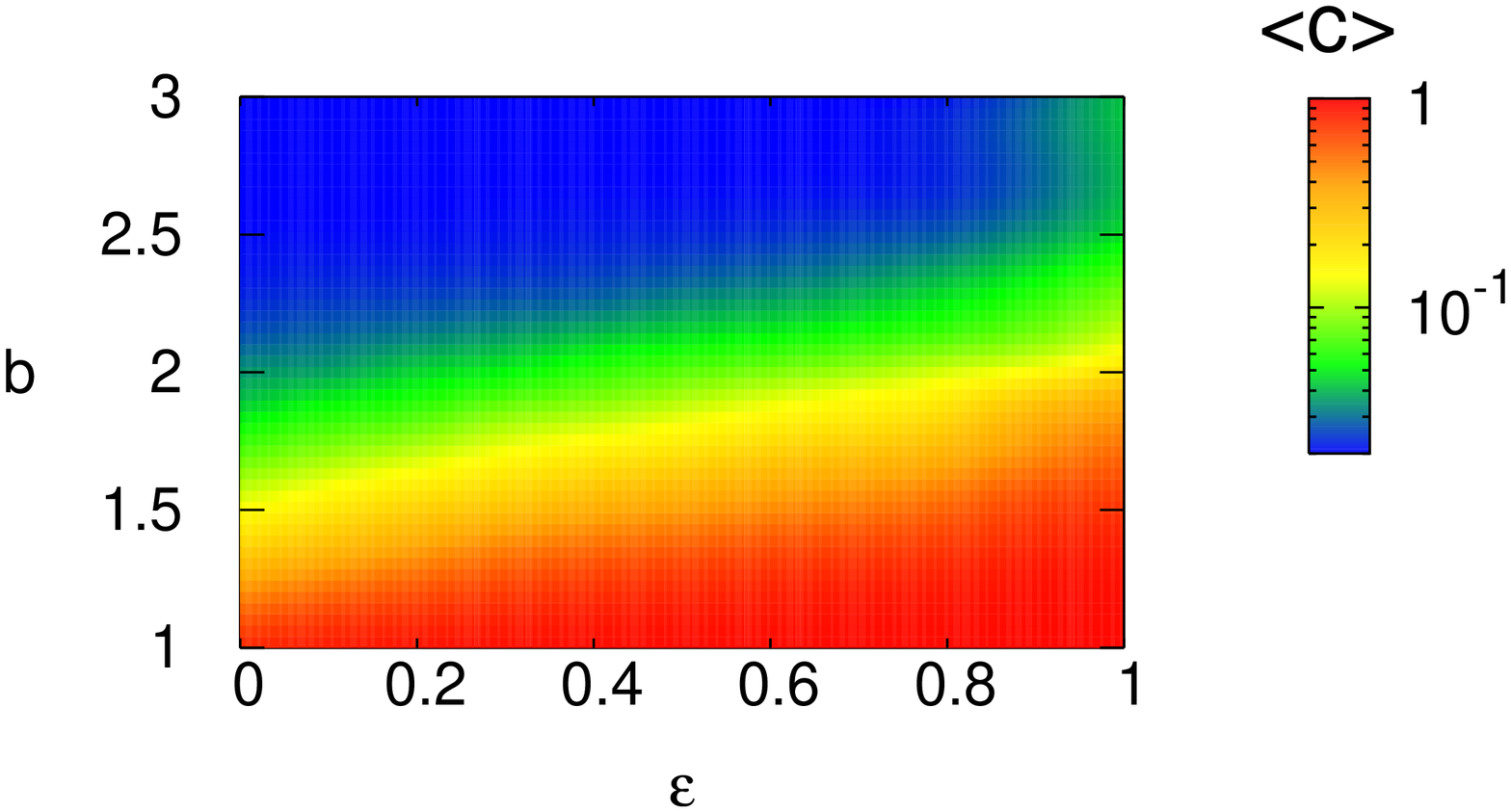,width=\columnwidth,angle=0,clip=1}
\end{center}
\caption{The upper panel shows degree distributions obtained for several values of $\epsilon$ for $b=1.5$. A transition from homogeneous to SF networks is evident. In the bottom panel, we have depicted the (color-coded) average level of cooperation, $\langle c\rangle$, as a function of the temptation to defect $b$ and the selection pressure $\epsilon$. The networks are made up of $10^3$ nodes with $\langle k \rangle=4$ and $\tau_D=10\tau_T$.}
\label{figure1}
\end{figure}

The growth of the network as defined above is thus linked to an evolutionary dynamics and controlled by the parameter $\epsilon$ and the two associated time scales ($\tau_{T}$ and $\tau_{D}$). When $\epsilon\simeq 0$, referred to as the weak selection limit \cite{Nowak1}, the network growth is independent of the evolutionary dynamics as all nodes are basically equiprobable. Alternatively, in the strong selection limit, $\epsilon\rightarrow 1$, the fittest players (highest payoffs) are 
much more likely to attract the newcomers. Therefore, Eq. (\ref{Pattach}) can be viewed as an ``{\em
Evolutionary Preferential Attachment}'' mechanism. We have carried out numerical simulations of the model exploring the ($\epsilon$, $b$)-space. In what follows, we focus on the results obtained when $\tau_{D}/\tau_{T}>1$, namely, the network growth is faster than the evolutionary dynamics \cite{note5}. Taking $\tau_{T}=1$ as the reference time, networks are generated by adding nodes every time step, while they play at discrete times given by $\tau_{D}$. As $\tau_{D}>\tau_{T}$, the linking procedure is done with the payoffs obtained the last time the nodes played \cite{note4}. All results for each value of $b$ and $\epsilon$ reported have been averaged over at least $10^3$ realizations and the 
number of links of a newcomer is taken to be $m=2$, whereas $m_0=3$.

\section{Results}

The dependence of the degree distribution on $\epsilon$ is shown in Fig.\ref{figure1} for $b=1.5$. As can be seen, the weak selection limit produces homogeneous networks characterized by a tail that decays exponentially fast with $k$. Alternatively, when $\epsilon$ is large, scale-free networks arise. Although this might a priori be expected from the definition of the growth rules, this needs not 
be the case: Indeed, it must be taken into account that in a one-shot PD game defection is the best
strategy regardless of the opponent strategy. However, if the network dynamics evolves into a state in which all players (or a large part of the network) are defectors, they will often play against themselves and their payoffs will be reduced. The system's dynamics will then end up in a state close to an all-$D$ configuration rendering $f_{i}(t)=0$ $\forall i$ $\in$ $[1,N(t)]$ in Eq.(\ref{Pattach}). From this point on, new nodes will attach randomly to other existing nodes [see Eq.(\ref{Pattach})] and therefore no hubs can come out. This turns out not to be the case, which indicates that for having some degree of heterogeneity, a nonzero level of cooperation is needed. Conversely, the heterogeneous character of the system provides a feedback mechanism for the survival of cooperators that would not outcompete defectors otherwise. 

\begin{figure}
\begin{center}
\epsfig{file=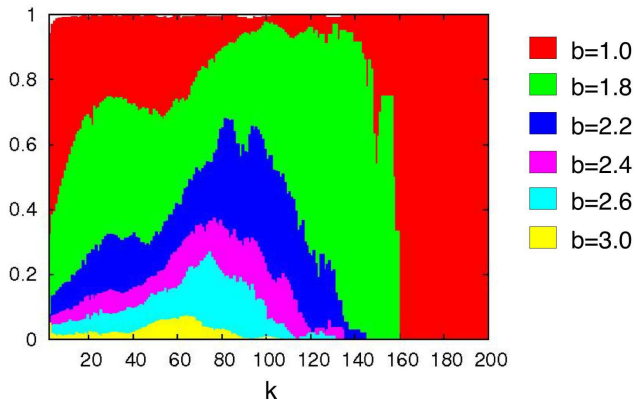,width=\columnwidth,angle=0,clip=1}
\end{center}
\caption{Probability that a node with connectivity $k$ plays as a cooperator for different values of $b$ in the strong selection limit ($\epsilon=0.99$) at the end of the growth of a network with $N=1000$ 
nodes.}
\label{figure2}
\end{figure}

The degree of heterogeneity of the networks in the strong selection limit depends slightly on $b$. The results indicate that when $\epsilon\rightarrow 1$, networks with the highest degree of heterogeneity, corresponding to the largest values of $b$, are not those with maximal cooperation levels. In Fig.\ \ref{figure1}, we have also represented the average level of cooperation, $\langle c\rangle$, as a function of the two model parameters $\epsilon$ and $b$. The figure shows that as $\epsilon$ grows for a fixed value of $b\gtrsim 1$, the level of cooperation increases. In particular, in the strong selection limit $\langle c\rangle$ attains its maximum value. This is a somewhat counterintuitive result as in the limit $\epsilon\rightarrow 1$, new nodes are preferentially linked to those with the highest payoffs, which for the PD game, should correspond to defectors. However, the population achieves the highest value of $\langle c\rangle$. On the other hand, higher levels of cooperation are achieved in heterogeneous rather than in homogeneous topologies, which is consistent with previous findings \cite{sp05,lhn05,gcfm07}. 

\begin{figure}
\begin{center}
\epsfig{file=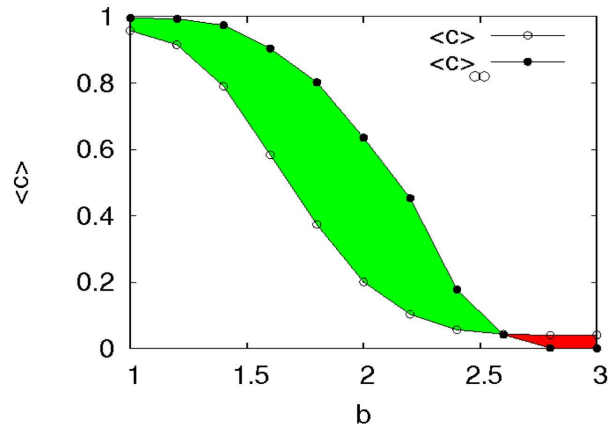,width=\columnwidth,angle=0,clip=1}
\end{center}
\caption{Degree of cooperation when the last node of the network is
    incorporated, $\langle c\rangle$, and the average fraction of cooperators
    observed when the system is time-evolved $\langle c\rangle_{\infty}$ after
    the network growth ends. Both magnitudes are shown as a function of $b$ for $\tau_D=10\tau_T$. See the text for further details.}
\label{figure3}
\end{figure}

The interplay between the local structure of the network and the hierarchical organization of cooperation is highly nontrivial. Contrary to what has been reported for static scale-free networks \cite{sp05,gcfm07},   Fig.\ref{figure2} shows that as the temptation to defect increases, the likelihood that cooperators occupy the hubs decreases. Indeed, during network growth, cooperators are localized neither at the hubs nor at the lowly connected nodes, but in intermediate degree classes. It is important to realize that this is a new effect that originates in the competition between network growth and the evolutionary dynamics. In particular, it highlights the differences between the microscopic organization in the steady state for the PD game in static networks with that found when the network is evolving. We will come back to this 
question in the Discussion section below. 

To confirm the robustness of the networks generated by evolutionary preferential attachment, 
let us consider the realistic situation that after incorporating a (possibly large) number of 
participants, network growth stops when a given size $N$ is reached, 
and that afterward only evolutionary dynamics takes place. In Fig. \ref{figure3}, we compare the average level of cooperation $\langle c\rangle$ when the network ceased to grow with the same quantity, but computed after allowing the evolutionary dynamics to evolve many more time steps $\langle c\rangle_{\infty}$ (without attaching new nodes). The green area indicates the region of the parameter $b$ where the level of cooperation increases with respect to that at the moment the network stops growing. On the contrary, the red zone shows that beyond a certain value of $b\approx 2.5$, cooperative behavior does not survive and the system dynamics evolves to an all-$D$ state. The increment of $\langle c\rangle$ when going from the steady state reached during network growth to the stationary regime attained once the underlying structure is static, has its roots on the fixation of cooperation in high degree classes, thus recovering the picture described in \cite{gcfm07}. On the other hand, when $b\gtrsim 2.5$, the few cooperators present in the growing network are not able to invade the hubs and finally, after a few more generations, cooperation is extinguished yielding $\langle c\rangle_{\infty}=0$. This result highlights the phenomenological difference between playing simultaneously to the growth of the underlying network and playing on fixed, static networks.

\begin{figure}
\begin{center}
\epsfig{file=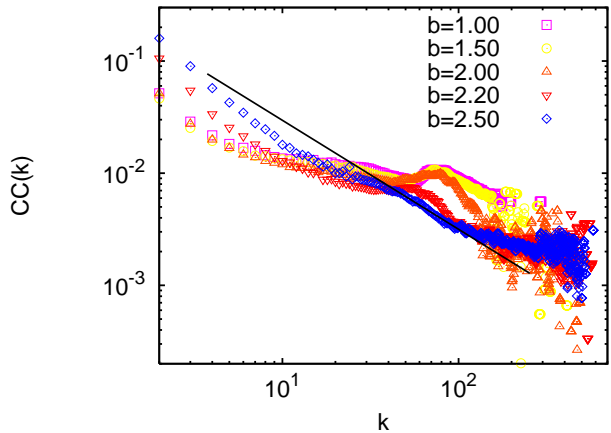,width=\columnwidth,angle=0,clip=1}
\end{center}
\caption{Dependence of the clustering coefficient $CC(k)\sim k^{-\beta}$ with the nodes' degrees for different values of $b$ in the strong selection limit. The straight line is a guide to the eye and corresponds to $k^{-1}$.}
\label{figure4}
\end{figure}

Another striking feature emerging from the interaction between network growth and the evolutionary dynamics is captured in Fig.\ \ref{figure4}, where the clustering coefficient, $CC$, has been represented as a function of the nodes degrees in the strong selection limit for several values of $b$. This coefficient measures the ratio of the number of triangles existing on the network over the total possible number of triangles, which relates to the possibility that a node connecting to a neighbor of another is also connected to this last one. Specifically, we will look at $CC(k)$, i.e., the way this coefficient depends on the degree of the node. Interestingly enough, the dependence of $CC(k)$ is consistent with a hierarchical organization expressed by the power law $CC(k)\sim k^{-\beta}$, a statistical feature found to describe many real-world networks \cite{PhysRep}. The behavior of $CC(k)$ in Fig.\ \ref{figure4} can be understood by recalling that in scale-free networks, cooperators are not extinguished even for large values of $b$ if they organize into clusters of cooperators that provide the group with a stable source of benefits \cite{gcfm07}. 

\section{Discussion}

Having presented our main simulation results, we now discuss them in detail and provide an 
interpretation of our observations that allows an understanding of the model behavior. To begin
with, let us consider the emergence of cooperation in the resulting network in the strong selection limit ($\epsilon\rightarrow 1$).
The organization of the cooperator nodes explains why cooperation survives and constitutes a unique positive feedback mechanism for the survival of cooperation. For simplicity, let us focus on how cycles of length $3$ (i.e., those contributing to $CC$) arise and grow. When a new node $j$ enters the network, it will preferentially attach to $m$ (recall we are using $m=2$) nodes with the highest payoff. Two situations are likely. On  the one hand, it may link to a defector hub with a high payoff. As the newcomer receives less payoff than the hub, it will sooner or later imitate its strategy and therefore will get trapped playing as a defector with $f_j=0$. Subsequently, node $j$ will not attract any links during network growth. On the other hand, if the new node attaches to a cooperator cluster, the other source of high payoff, and forms a triad with the cluster elements, two outcomes are possible depending on its initial strategy. If the newly attached node plays as a defector, the triad may eventually be invaded by defectors and may end up in the long run in a state where the nodes have no capacity to receive new links. Conversely, if it plays as a cooperator, the group will be reinforced, both in its robustness against defector invasion and in its overall fitness to attract new links, i.e., playing as a cooperator while taking part in a successful (high fitness) cooperator cluster reinforces its future success, while playing as a defector undermines its future fitness and leads to dynamically (and topologically) frozen ($f_i=0$) structures, so that defection cannot take long-term advantage from cooperator clusters. Therefore, cooperator clusters that emerge from cooperator triads to which new cooperators are attached can then continue to grow if more cooperators are attracted or even if defectors attach to the nodes whose connectivity verifies $k>mb$. Moreover, the stability of cooperator clusters and its global fitness grow with their size, specially for their members with higher degree, and naturally favors the formation of triads among its components. 
Note, additionally, that it follows from the above mechanism that a node of degree $k$ is a vertex of $(k-1)$ triangles and then $CC(k)=\frac{(k-1)}{k(k-1)/2}=2/k$, the sort of functional form for the clustering coefficient reported in Fig.\ \ref{figure4}. 

Another interesting phenomenon arising from our model is the fact, previously
unobserved, that cooperators occupy the nodes with intermediate degree and the hubs are defectors, 
in contrast with the simulations on static networks \cite{gcfm07,pgfm07}. To address this issue we have developed a simple analytical argument. Let
 $k_i^c$ be the number of cooperator neighbors of a given node $i$.
Its fitness is $f_i^d=bk_i^c$, if node $i$ is a defector, and
$f_i^c=k_i^c$, if it is a cooperator. The value of $k^c_i$ is
expected to change due to both network growth (node accretion
flow, at a pace of one new node each time unit $\tau_T$) and 
imitation processes that take place at a pace
$\tau_D$. We will focus on the case in which $\tau_D$
is much larger than $\tau_T$. The expected increase of fitness is
\begin{equation}
\Delta f_i = \Delta_{flow} f_i + \Delta_{evol} f_i,
\label{nueva1}
\end{equation}
where $\Delta_{flow}$ means the variation of fitness in node $i$ due to the newcomers flow, and $\Delta_{evol}$ stands for the change in fitness due to changes of neighbors' strategies. The above expression would lead to an expected increase in $k^c_i$ given by 
\begin{equation}
k^c_i(t+\tau_D)- k^c_i(t)=\Delta k^c_i = \Delta_{flow} k^c_i + \Delta_{evol} k^c_i.
\label{nueva2}
\end{equation}
On the other hand, the expected increase of degree in the interval $(t,t+\tau_D)$ only has the contribution from newcomers flow and takes the form (recall that
new nodes are generated with the same probability to be cooperators
or defectors)
\begin{equation}
\Delta k_i = \Delta_{flow} k_i = 2 \Delta_{flow} k^c_i.
\label{nueva3}
\end{equation}

If the fitness (hence connectivity) of node $i$ is high enough as to
attract a significant part of the newcomers flow, the first term in Eq. (\ref{nueva1})
dominates
at short time scales, and then the hub degree $k_i$ increases
exponentially. Connectivity patterns are then dominated by the
growth by preferential attachment, ensuring as in the Barab\'asi-Albert \cite{bara}
model that the network
will have a SF degree distribution. Moreover, the rate of increase
\begin{equation}
\Delta_{flow} k^c_i=\frac{1}{2} m \tau_D \frac{f_i}{\sum_j f_j}
\label{nueva4}
\end{equation}
is larger for a defector hub (by a factor $b$) because of its larger fitness, 
and then one should expect hubs to be
mostly defectors, as confirmed by the results shown in Fig.\ \ref{figure2}.
This small set of most connected defector nodes attracts most of
the newcomers flow. 

On the contrary, for nodes of intermediate degree, say of
connectivity $m \ll k_i \ll k_{max}$, the term $\Delta_{flow}f_i$
in Eq.\ (\ref{nueva1})
can be neglected, i.e., the arrival of new nodes is a rare
event, so that for a large time scale, $\dot{k}_i=0$. Note that if $\dot{k}_i(t)=0$ for all $t$ in an interval $t_0\leq t \leq t_0+T$, the size of the neighborhood is constant during the whole interval $T$ and thus the evolutionary dynamics of strategies through imitiation 
is the exclusive responsible for the
strategic field configuration in the neighborhood of node $i$. During
these stasis periods the probability distribution of strategies approaches that of a
static network in the neighborhood of node $i$. 
It is clear that this scenario can be occasionally
subject to sudden (avalanche-type of) perturbations
following "punctuated equilibrium" patterns in the rare ocasions in
which a new node arrive. Recalling that the probability for
this node $i$ of intermediate degree  to be a 
cooperator is large in the static regime \cite{gcfm07} we then 
arrive to the conclusion that for these nodes the density of 
cooperators must reach a maximum, in agreement with 
Fig.\ \ref{figure2}. Furthermore, our simulations show that 
these features of the shape of the curve are indeed preserved as time goes by, giving further support to the above argument based on time scale separation and confirming that our understanding
of the mechamisms at work in the model is correct.

\section{Conclusions}

In summary, we have presented a model in which the rules governing the formation of the network are linked to the dynamics of its components. The model provides an evolutionary explanation for the origin of the two most common types of networks found in natural systems: When the selection pressure is weak, homogeneous networks  arise, whereas strong selection pressure gives rise to scale-free networks. A remarkable fact is that the proposed evolution rule gives rise to complex networks that share many topological features with those measured in real systems, such as the power law dependence of the clustering coefficient with the degree of the nodes. Interestingly, our results make it clear that 
the microscopic dynamical organization of strategists in evolutionarily grown networks is very 
different from the case in which the population evolves on static networks. Furthermore, as we have
seen, the generated networks are robust in the sense that after the growth process stops, the 
dynamical behavior keeps its character. 

Thinking of the specific application 
we are discussing here, the emergence of cooperation, it is particularly remarkable the special role 
of individuals with an intermediate number of connections. As we have reasoned above, as time
proceeds and the network grows, cooperation increases by invading those intermediate nodes, 
and on the other hand the range of intermediate degrees grows as well, leading to further increase
of cooperation. On the contrary, hubs or well connected nodes, which on the static scenario are the
supporters of cooperation, in the evolutionary process are defectors that thrive and accumulate new
nodes by being so, only to fall eventually in the class of intermediate degree nodes and become 
cooperators. The analogy with the effect of a well-doing middle class in a western-like society is tempting but would of course
be too far-fetched to push it beyond a general resemblance. Nevertheless, one particular situation in which models like this, based on 
the evolutionary preferential attachment mechanism, may prove
very relevant is in the formation of social networks of entrepreneurs
or professionals, such as those studied in Silicon Valley 
\cite{saxenian,alex}. The way these networks grow upon arrival of
new individuals and subsequent cooperative interactions 
made them a natural scenario to apply these ideas in detail. Finally, another important conclusion is the resilience
of the cooperative behavior arising in these networks, in so far as it does not decrease for a wide
range of parameters upon stopping the growth process, and, in most cases it even exhibits a 
large increase of the cooperation level. 

On more general theoretical grounds, figuring out why scale-free networks are so ubiquitous in Nature is one of the most challenging aspects of modern network theory. At variance with previous hypotheses, the evolutionary preferential attachment mechanism of Eq. (\ref{Pattach}) naturally incorporates a competition between structural and dynamical patterns and hence it suffices to explain why SF networks are optimized to show both structural and dynamical robustness. The former is given by the scale-free nature of the resulting topology, while the latter is based on the high levels of cooperation attained in the grown networks. Note that this optimization acts at a local level since individuals search  their own benefit rather than following a global optimization scheme \cite{doneti}, to be compared with the 
fact that the 
resulting network has a very good cooperation level as a whole.  Finally, we let for future research the question of whether Eq. (\ref{Pattach}) can be applied to other sort of dynamics by appropriately defining the dynamical variable $f_i(t)$ and adjusting the growth rules. It is however reasonable to assume that the functional form in Eq. (\ref{Pattach}) may render general for generating optimized SF networks.

\begin{acknowledgments}
We acknowledge support from the Ministerio de Educaci\'on y Ciencia
through the Ram\'on y Cajal Program (Y.M.) and grants FIS-2006-12781-C02-01,
FIS-2005-00337, MOSAICO and NAN2004-9087-C03-03. A.S. is also 
supported by the Comunidad de Madrid (Spain) under grant SIMUMAT-CM.
\end{acknowledgments}


\begin{thebibliography}{99}
\bibitem{note3} The only exception is the initial core which is a fully connected cooperator network.
\bibitem{note2} We have also explored other forms for the probability Eq.\ (\ref{replicator}) such as unconditional imitation, with no significant qualitative changes.
\bibitem{note5} The dependence of the network properties and average level of cooperation on $\tau_{D}/\tau_{T}>1$ is weak. We have checked that using $\tau_{D}/\tau_{T}=1,5,10, \text{and } 20$  produces the same qualitative results. On the other hand, the choice of $\tau_{D}/\tau_{T}<1$ seems not to be realistic as selection should be slower than growth. In these cases, cooperation and defection can not coexist as the system dynamics evolves either to an all-C or to all-D configuration. However, networks with diverse degree of heterogeneity can be generated. 
\bibitem{note4} The payoffs are not accumulated, i.e., before playing, $f_{i}(t)$ is set to zero $\forall i$ $\in$ $[1,N(t)]$.
\end{thebibliography}
\end{document}